\documentclass[times, 10pt,twocolumn]{article} 
\usepackage{latex8}
\usepackage{times}

\usepackage{color}
\usepackage{graphicx}
\usepackage{algorithm}
\usepackage{algorithmic}
\usepackage{amsthm}
\usepackage{amsmath}
\usepackage{subfigure}
\usepackage{geometry}

\pdfoutput=1

\newcommand{\rem}{\mathrm{rem}}
\newcommand{\G}{\mathrm{G}}
\newcommand{\W}{\mathrm{W}}
\newcommand{\T}{\mathrm{T}}
\newcommand{\RT}{\mathrm{RT}}

\newcommand{\DASH}{\mathrm{DASH}}
\newcommand{\SDASH}{\mathrm{SDASH}}

\title{Picking up the Pieces: Self-Healing in Reconfigurable Networks}
{\author{Jared Saia \thanks{Department of Computer Science,  University of New
Mexico,  Albuquerque,  NM 87131-1386; email: \{saia, amitabh\}@cs.unm.edu. This research was partially supported by NSF CAREER Award 0644058, NSF CCR-0313160, and an AFOSR MURI grant.}
\and Amitabh Trehan
 \footnotemark[1]
}}

\newtheorem{lemma}{Lemma}

\newtheorem{theorem}{Theorem}

\DeclareGraphicsExtensions{.pdf}

\begin{document}
\date{}
\maketitle

\begin{abstract}

We consider the problem of self-healing in networks that are \emph{reconfigurable} in the sense that they can change
their topology during an attack.  Our goal is to maintain connectivity in these networks, even in the presence of
repeated adversarial node deletion, by carefully adding edges after each attack.  We present a new algorithm, $\DASH$, that
provably ensures that: 1) the network stays connected even if an adversary deletes up to all nodes in the network; and
2)  no node ever increases its degree by more than $2 \log n$, where $n$ is the number of nodes initially in the
network.  $\DASH$ is fully distributed; adds new edges only among neighbors of deleted nodes; and has average latency and
bandwidth costs that are at most logarithmic in $n$.   $\DASH$ has these properties irrespective of the topology of the
initial network, and is thus orthogonal and complementary to traditional topology-based approaches to defending against
attack. 

We also prove lower-bounds showing that $\DASH$ is asymptotically optimal in terms of minimizing maximum degree increase
over multiple attacks.  Finally, we present empirical results on power-law graphs that show that $\DASH$ performs well in
practice, and that it significantly outperforms naive algorithms in reducing maximum degree increase.

\end{abstract}

\Section{Introduction}

On August 15, 2007 the Skype network crashed for about $48$ hours, disrupting service to approximately $200$ million users~\cite{fisher,malik,
moore, ray, stone}.  Skype attributed this outage to failures in their ``self-healing mechanisms''~\cite{garvey}.  We believe
that this outage is indicative of a much broader problem.
Modern computer systems have complexity unprecedented in the history of engineering: we are approaching scales of billions of components.  Such systems are less akin to a traditional engineering enterprise such as a bridge, and more akin to a living organism in terms of complexity.  A bridge must be designed so that key components never fail, since there is no way for the bridge to automatically recover from system failure.  In contrast, a living organism can not be designed so that no component ever fails: there are simply too many components.  For example, skin can be cut and still heal. Designing skin that can heal is much more practical than designing skin that is completely impervious to attack.  Unfortunately, current algorithms ensure robustness in computer networks through hardening individual components or, at best, adding lots of redundant components.  Such an approach is increasingly unscalable.

In this paper, we focus on a new, \emph{responsive} approach for maintaining robust networks.  Our approach is
responsive in the sense that it responds to an attack (or component failure) by changing the topology of the network. 
Our approach works irrespective of the initial state of the network, and is thus orthogonal and complementary to
traditional  non-responsive techniques.  There are many desirable invariants to maintain in the face of an attack.  Here
we focus only on one of the simplest and most fundamental invariants: maintaining network connectivity.

The responsive approach will only work on networks that are \emph{reconfigurable}, in the sense that the topology of the
network can be changed.  Not all networks have this property.  However, many large-scale networks are reconfigurable. 
For example, peer-to-peer and overlay networks are reconfigurable, as are wireless and mobile networks.  More generally,
many social networks, such as a company's organizational chart;  infrastructure networks, such as an airline's
transportation network; and biological networks, such as the human brain, are also reconfigurable.  The increasing
importance of these types of networks calls for new mathematical and algorithmic methods to study and exploit their
flexibility.

\medskip
\noindent
{\bf Our Model:} We now describe our model of attack and network response.  We assume that the network is initially a
connected graph over $n$ nodes.  We assume that every node knows not only its neighbors in the network but also the
neighbors of its neighbors i.e. neighbor-of-neighbor (NoN) information.  In particular, for all nodes $x$,$y$ and $z$
such that $x$ is a neighbor of $y$ and $y$ is a neighbor of $z$, $x$ knows $z$.  There are many ways that such
information can be efficiently maintained, see e.g.~\cite{MNW,NW}.

We assume that there is an adversary that is attacking the network.  This adversary knows the network topology and our
algorithm, and it has the ability to delete carefully selected nodes from the network.  However, we assume the adversary
is constrained in that in any time step it can only delete a small number of nodes from the network\footnote{Throughout
this paper, for ease of exposition, we will assume that the adversary deletes only one node from the network before the
algorithm responds.  However, our main algorithm, $\DASH$, can easily handle the situation where any number of nodes are
removed, so long as the neighbor-of-neighbor graph remains connected.}.  We further assume that after the adversary
deletes some node $x$ from the network, that the neighbors of $x$ become aware of this deletion and that they have a
small amount of time to react.

When a node $x$ is deleted, we allow the neighbors of $x$ to react to this deletion by adding some set of edges amongst
themselves.  We assume that these edges can only be between nodes which were previously neighbors of $x$.  This is to
ensure that, as much as possible, edges are added which respect locality information in the underlying network.  We
assume that there is very limited time to react to deletion of $x$ before the adversary deletes another node.  Thus, the
algorithm for deciding which edges to add between the neighbors of $x$ must be fast and localized.

\noindent {\bf Our Results:}  We introduce an algorithm for self-healing of reconfigurable networks, called  $\DASH$
(an acronym for \emph{Degree based Self-Healing}). $\DASH$ is \emph{locality-aware} in that it uses only the
neighbors of the deleted node for reconnection.  We prove that $\DASH$ maintains connectivity in the network, and
that it increases the degree of any node by no more than $O(log n)$.  During reconnection of nodes, our algorithm uses
only local information, therefore, it is scalable and can be implemented in a completely distributed manner. Algorithm
 $\DASH$ is described as Algorithm \ref{algo: dash} in Section~\ref{sec: dash}.  The main characteristics of $\DASH$ are
summarized in the following theorem that is proved in Section~\ref{sec: dash}.

\begin{theorem} 
$\DASH$ guarantees the following properties even if up to all the nodes in the network are deleted:
\begin{itemize}
\item The degree of any vertex is increased by at most $2 \log n $.
\item The number of messages any node of initial degree $d$ sends out and receives is no more than $2 (d + 2 \log n) \ln
n $
\emph{with high
probability}\footnote{Throughout this paper, we use the phrase with high probability (w.h.p) to mean with probability at
least $1-1/n^{C}$ for any fixed constant $C$.} over all node deletions.
\item The latency to reconnect is $O(1)$ after attack; and the amortized latency to update the state of the network over $\theta(n)$ deletions 
 is $O(\log n)$ with high probability. 
\end{itemize}
\end{theorem}

\noindent
We also prove (in Section~\ref{sec: lower}) the following lower bound that shows that  $\DASH$ is
asymptotically optimal.

\begin{theorem}
 Consider any locality-aware algorithm that increases the degree of any node after an attack by at most a
fixed constant.  Then there exists a graph and a strategy of deletions on that graph that will force the algorithm to
increase the degree of some node  by at least $\log n$.
\end{theorem}

We also present empirical results (in Section~\ref{sec: empirical}) showing that $\DASH$ performs well in practice and that
it significantly outperforms naive algorithms in terms of reducing the maximum degree increase. Finally (in Section~\ref{sec: empirical}) we describe $\SDASH$, a heuristic based on $\DASH$ that we show empirically both keeps node degrees small and also keeps shortest paths between nodes short.

\medskip
\noindent
{\bf Related Work:}  There have been numerous papers that discuss strategies for adding additional  
capacity and rerouting in anticipation of failures \cite{ doverspike94capacity, frisanco97capacity, iraschko98capacity, 
 murakami97comparative, caenegem97capacity, xiong99restore}. 
Here we focus on results that are responsive in some sense.  M\'{e}dard, Finn, Barry, and Gallager \cite{medard99redundant} propose constructing redundant trees to make backup routes possible when an edge or node is deleted.  Anderson, Balakrishnan, Kaashoek, and Morris \cite{anderson01RON} modify some existing nodes to be RON (Resilient Overlay Network) nodes to detect failures and reroute accordingly. Some networks have enough redundancy built in so that separate parts of the network can function on their own in case of an attack~\cite{goel04resilient}.  In all these past results, the network topology is fixed.  In contrast, our algorithm adds edges to the network as node failures occur.  Further, our algorithm does not dictate routing paths or specifically require redundant components to be placed in the network initially.  In this paper, we build on earlier work done in \cite{BomanSAS06, IchingThesis}, which proposed a simple line algorithm for self-healing to maintain network connectivity.

\medskip
\noindent
{\bf Table of Contents:}  
The rest of our paper is organized as follows.  Section~\ref{sec: dash} describes the algorithm $\DASH$, and its
theoretical properties.  Section~\ref{sec: lower} gives a lower bound on locality-aware algorithms.  Section~\ref{sec: empirical} gives empirical results for $\DASH$, and several other simple
algorithms on random power-law networks. It also describes and gives results for $\SDASH$. We conclude and give areas for future work in Section~\ref{sec: conclusions}. 


\Section{$\DASH$: An Algorithm for Self-Healing}
\label{sec: dash}
\setcounter{theorem}{0}

In this Section, we describe $\DASH$ and prove certain properties about it. In brief, when a deletion occurs, $\DASH$ asks the neighbors of the deleted node to reconnect themselves into a certain kind of complete binary tree. Then messages are propagated so that the nodes can keep track of which connected component they belong to. 

Let the actual network at a particular time step be $G(V,E)$. Let $E'$ be the edges (i.e. \emph{healing edges}), that have been added by the algorithm up to that time step (note $E'\subseteq E$). 
 Let $G'=(V, E')$. We show that $G'$ is a forest in Lemma \ref{lemma: forest}.

\SubSection{$\DASH$: Degree based Self-Healing}
\label{subsec: dash}

 As the acronym suggests, $\DASH$ employs information of previous degree increase to control further degree increase for a node. When a deletion occurs, we assume the neighbors of the deleted node are able to detect the deletion. Then they employ $\DASH$ to heal. To maintain connectivity, $\DASH$ connects the neighbors of a deleted node as a binary tree. The tree is structured so that the vertices which have incurred the maximum degree increase previously get to be leaves and thus not increase their degree in this round. Notice that at least half the vertices in a binary tree are leaves. The nodes maintain information about the virtual network and their connected component in this network. The algorithm tries to use only a single node from each component during reconnection and thus adds only a low number of new edges during healing.

 To describe $\DASH$ we give some definitions. Let $N(v,G)$ be the neighbors of vertex $v$ in the graph $G$ representing
the real network.  Let $N(v,G')$ be the neighbors of vertex $v$ in graph $G'$ consisting of the edges added by the
healing algorithm. Let $\delta(v)$ be the degree increase of the vertex $v$ compared to its initial degree. Note that
this is not the same as the degree of  $v$ in $G'$. 

When a node $v$ is deleted, 
partition  on the basis of their $ID$ all the neighbors of $v$ in $G$ (not having the same $ID$ as $v$). Let $UN(v,G)$
(\emph{Unique Neighbors})
be  the  set having one representative from each of the partitions. If there is  more than one node as a  possible
representative from a partition, we include the one with the lowest initial $ID$.

 Note that $UN(v,G) \cap N(v,G') = \phi$ and $UN(v,G) \cup N(v,G') \subseteq N(v,G)$ . The $ID$
of a node allows us to keep track of which connected component in $G'$ it belongs to.  The lowest $ID$ of any node in
that component is broadcast and  all the nodes in the component take on this $ID$. 

\begin{algorithm}[h!]
\caption{\textbf{DASH:} Degree-Based Self-Healing}
\label{algo: dash}
\begin{algorithmic}[1]
\STATE \emph{Init:} for given network $G(V,E)$, Initialize each vertex with a random number $ID$ between [0,1] selected
uniformly at random. 
\WHILE {true}
\STATE \emph{If a vertex $v$ is deleted, do}
\STATE Nodes in $UN(v,G) \cup N(v,G')$ are reconnected into a \emph{complete binary tree}. To connect the tree, go left
to right, top down, mapping nodes to the \emph{complete binary tree} in increasing order of $\delta$ value.

\STATE Let $MINID$ be the minimum $ID$ of any node in $UN(v,G) \cup N(v,G')$.
 Propagate $MINID$ to all the nodes in the tree of $UN(v,G) \cup N(v,G')$ in $G'$. All these nodes now set their $ID$ to
$MINID$.
\ENDWHILE
\end{algorithmic}
\end{algorithm}

Our main results about $\DASH$ are stated in Theorem~\ref{theorem: DASH}.

\begin{theorem} 
$\DASH$ is a distributed algorithm with the following properties:
\label{theorem: DASH}
\begin{itemize}
\item The degree of any vertex is increased by at most $2 \log n $.
\item The latency to reconnect is $O(1)$.
\item The number of messages any node of degree $d$ sends out and receives is no more than $(2d + 2 \log n) \ln n $ \emph{with high probability} over all node deletions.
\item The amortized latency for  $ID$ propagation is $O(log n)$ \emph{with high probability} over all node deletions. 
\end{itemize}
\end{theorem}

\SubSection{Proof of Theorem~\ref{theorem: DASH}}

For analysis, we use the following definitions:  
\def \max{\mathop{\rm max}\limits}%
\begin{itemize}
   \item Let $T(x,y)$ be the tree in $G'-y$ that contains $x$.     
\item Each vertex $v$ will have a weight, $w(v)$. The weight of a vertex will start at 1 and may increase during the
algorithm. If $v$ is deleted, $w(v)$ is added to an arbitrarily chosen neighbor in $G'$.
     \item Let $W(S) = \sum\limits_{v \in V} w(v)$, for a graph $S(V,E)$ i.e. the sum of the weights of all vertices in $S$.
   \item For vertex $v$, let $\rem(v)$ =
 \[
\sum_{u \in N(v,G')}\!\!\!\!\!\! W(T(u,v))\, - 
  \max_{u \in N(v,G')}\!\!\!\!\!\!( W(T(u,v)))\, + w(v).
\]
             We will show that as the degree of a vertex increases in our algorithm, so will the  $\rem$ value of that vertex. 
           Intuitively $\rem(v)$ is large when removing $v$ from its tree in $G'$ gives rise to many connected components with large weight.

\end{itemize}

\begin{lemma}
\label{lemma: forest} The edges added by the algorithm, $E'$, form a forest.
\end{lemma}
\begin{proof}We prove this by induction on the number of nodes deleted.\\

\noindent\emph{Base Case:} Initially, $G'$ is a forest because $E'$ is empty.\\

\noindent We note that $E'$ and $G'$ change only when a deletion occurs. Consider the $i^{th}$ deletion and let $v$ be
the node deleted.\\
 Let $v$ belong to tree $T_v$ in $G'$ just prior to the deletion of $v$.
Now, for all $ x, y  \in N(v,G')$\ x and y are not connected in $E'$ since that would have implied the existence of a
cycle through $v$ contradicting the Inductive Hypothesis. Note also that for all $z \in UN(v,G), z \notin T_v$. Since we
select only 1 node from each tree $T_i$ in which $v$ had a neighbor, no pair of nodes in $UN(v,G) \cup N(v,G')$ are
connected in $G'$. We reconnect all the nodes in $UN(v,G) \cup N(v,G')$ in a Binary Tree and propagate the minimum ID.
Since we are adding edges between nodes which previously were in separate connected components in $G'$, no cycles are
introduced. Hence, $G'$ remains a forest. 

\end{proof}

\begin{lemma}
\label{lemma: nondecreasing} For any vertex $v$, $rem(v)$ is non-decreasing over any vertex deletion where $v$ has not
been deleted.
\end{lemma}
\begin{proof}

By Lemma \ref{lemma: forest}, every vertex $v$ in $G'$ belongs to some tree, which we will call $T_v$. For every $T_v$
in $G'$,  $W(T_v)$ is the sum of the weights 
of all vertices in $T_v$.

By definition,
$\rem(v)$ =
 \[
\sum_{u \in N(v,G')}\!\!\!\!\!\! W(T(u,v))\, - 
  \max_{u \in N(v,G')}\!\!\!\!\!\!( W(T(u,v)))\, + w(v).
\]


Therefore,\\
 $rem(v) =  W(T_v) - \max_{u \in N(v,G')} W(T(u,v))$

Observe first that $W(T_v)$ cannot decrease even when there is a deletion in $T_v$ because the deleted vertex's weight
is not ``lost", but added to some member of $T_v$. 
 
 
 Since $W(T_v)$ cannot decrease, $rem(v)$ can only 
decrease if the maximum subtree weight increases more than $W(T_v)$. 
Since the maximum subtree is a subset of the tree, $T_v$, 
any increases or decreases in the maximum subtree is also counted in $W(T_v)$. Thus,  $rem(v)$ cannot decrease.

\end{proof}

\begin{lemma}
\label{lemma: WgreatRem}
 For any node $v$, for all nodes $q \in N(v,G')$ , $W(T(v,q)) \ge rem(v)$.
 \end{lemma}

\begin{figure}[h!]
\centering
\includegraphics[scale=0.5]{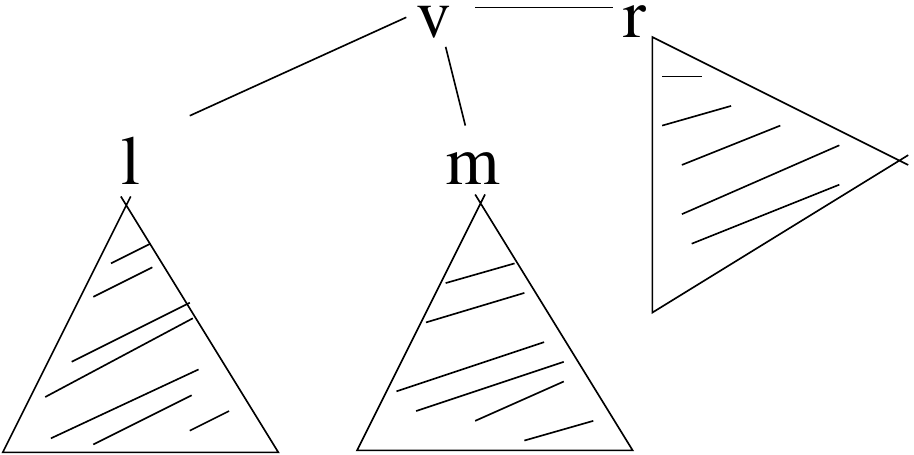}
\caption{  $W(T(v,m)) \ge rem(v)$.}
\label{fig: WgreatRem}
\end{figure}

\begin{proof}
\noindent For all nodes $q$,
\begin{eqnarray*}
   W(T(v,q)) & = &\sum\limits_{u \in N(v,G')\atop{u \neq q}} W(T(u,v)) + w(v)\\
    & \ge & \sum\limits_{u \in N(v,G')} W(T(u,v))\\
     & & - \max_{u \in N(v,G')} W(T(u,v)) + w(v)\\ 
    & = & rem(v)
 \end{eqnarray*}
   

For example, in figure \ref{fig: WgreatRem}, $W(T(V,M)) = W(T(L,V)) + W(T(R,V)) + w(v) \ge rem(v)$.
\end{proof}

\begin{lemma}
\label{lemma: deglimit}
  For any node v, $rem(v) \ge 2^{\delta(v)/2}$, where $\delta(v)$, as defined earlier, is the degree increase of the
vertex $v$ in $G$.
\end{lemma}

\begin{proof} Let $t$ be the number of rounds of healing where a round is a single adversarial deletion followed by
self-healing by $\DASH$. We prove this lemma by induction on $t$.\\
 Let $\G'_{t}$, $\rem_t(v)$ and $\delta_{t}(v)$ be $\G'$, $\rem(v)$ and $\delta(v)$ respectively at time $t$.\\

\emph{Base Case:} t = 0:
  In this case, all nodes $v$  have  $\delta(v) = 0$; $\rem(v) = 1$.  Thus, $\rem(v) \ge 2^{0}$.\\

 \emph{Inductive Step:}
 Consider the network at round $t$. We assume by the inductive hypothesis that for all nodes $v$ in $\G'$,
$\rem_{t-1}(v) \ge 2^{\delta_{t-1}(v)/2} $. Our goal is to  show that $\rem_{t}(v) \ge 2^{\delta_{t}(v)/2} $.

 Suppose node $x$ was deleted at round $t$. According to our algorithm, some or all of the neighbors of $x$ will be
reconnected as a binary tree. Let us call this  tree $\RT$ (short for \emph{Reconstruction Tree}).  Let $T(x,y)$ be
the tree in $G'_{t-1}-y$ that contains $x$, and $T'(x,y)$ be the tree in $G'_{t}-y$ that contains $x$.

 Consider a surviving vertex $v$. If $v$ is not a part of $\RT$, then by a simple application of lemma \ref{lemma:
nondecreasing}, our induction holds. If $v$ is a part of $\RT$, there are 3 possibilities:

\begin{enumerate}
\item \emph{$v$ is a leaf node in $\RT$}

 The degree of $v$ did not change. Thus, $\delta_t(v) = \delta_{t-1}(v)$. By Lemma  \ref{lemma: nondecreasing},
$\rem_{t}(v) \ge \rem_{t-1}(v)$. Thus, using  the induction hypothesis, $\rem_{t}(v) \ge 2^{\delta_{t}(v)/2} $.

\item \emph{$v$ is the root of $\RT$}

\begin{figure}[h!]
\centering
\includegraphics[scale=0.3]{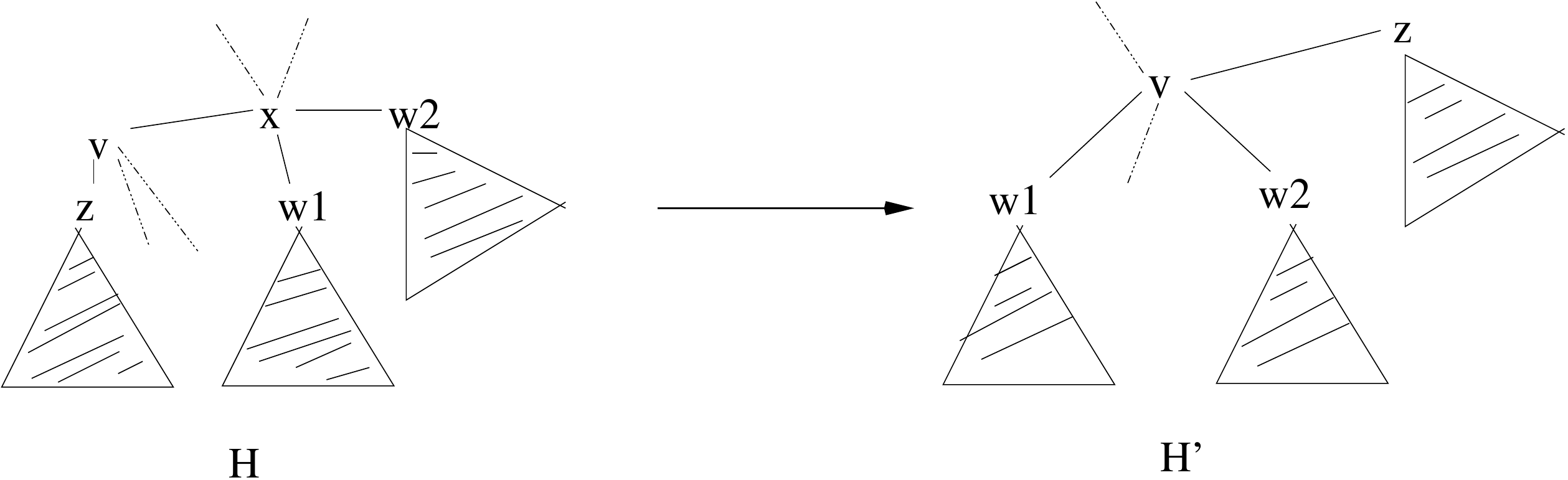}
\caption{node $v$ is the root, with 2 children}
\label{fig: vroottree}
\end{figure}

 If $v$ has only one child in $\RT$, then this is the same as the previous case with the parent and child role reversed
and the induction holds. Let us consider the case when $v$ has two children in $\RT$. Now, $\delta_t(v)$ has increased
by 1. Let $z$ be the neighbor of $v$ such that $W(T(z,v))$ is the largest among all neighbors of $v$ except $x$. Note
that $W(T'(z,v)) = W(T(z,v))$, since this subtree was not involved in the reconstruction.
Consider the possibly empty subtree of $v$ rooted at $z$.
  Let the two children of $v$ in $\RT$ be $w_1$ and $w_2$, as illustrated in figure \ref{fig: vroottree}. By our
algorithm, we
know that $\delta_{t-1}(w_1) \ge \delta_{t-1}(v)$ and $\delta_{t-1}(w_2) \ge \delta_{t-1}(v)$. Thus, using the inductive
hypothesis and lemma
\ref{lemma: WgreatRem}, we have that $\W(\T(w_1,x)) \ge \rem_{t-1}(w_{1}) \ge 2^{\delta_{t-1}(w_{1})/2}$ and $\W(\T(w_2,x))
\ge \rem_{t-1}(w_{2}) \ge 2^{\delta_{t-1}(w_{2})/2}$. 
 By lemma \ref{lemma: nondecreasing}, this implies that in $G'_{t}$,
  \begin{eqnarray*}
    W(\T'(w_{1},v)) & \ge  2^{\delta_{t-1}(w_{1})/2} & \ge 2^{\delta_{t-1}(v)/2}\\
    W(\T'(w_{2},v)) & \ge  2^{\delta_{t-1}(w_{2})/2} & \ge 2^{\delta_{t-1}(v)/2}
   \end{eqnarray*}
   
  Assume without loss of generality that   $\W(\T'(w_1,v)) \le \W(\T'(w_2,v))$. There are two cases:

\begin{enumerate}
\item $\W(\T(z,v)) < \W(\T'(w_1,v))$

  In this case $\rem_{t-1}(v)$ did not include $\W(\T(x,v))$.
 But  $\rem_t(v)$ will include $\W(\T'(w_1,v))$
 Hence,\\
\begin{eqnarray*}
 \rem_t(v) & \ge & \rem_{t-1}(v) + \W(\T'(w_1,v))\\
  & \ge & 2^{\delta_{t-1}(v)/2} + 2^{\delta_{t-1}(v)/2}\\
  & = & 2^{(\delta_{t-1}(v)+2)/2} \\
  & = & 2^{(\delta_{t}(v)+1)/2}
  \end{eqnarray*}

\item $\W(\T(z,v)) \ge \W(\T'(w_1,v))$

  In this case  $\rem_t(v)$ will include $\W(\T'(w_1,v))$ and the smaller of  $\W(\T'(w_2,v))$ and  $\W(\T'(z,v))$.
  Note that by Lemmas \ref{lemma: WgreatRem} and \ref{lemma: nondecreasing}, the inductive hypothesis, and the fact
that $\delta_{t-1}(w_{1})
\ge
\delta_{t-1}(v) $,  $\W(T'(w_1,v)) \ge \rem_{t}(w_{1}) \ge \rem_{t}(w_{1}) \ge 2^{\delta_{t-1}(w_{1})/2} \ge 
2^{\delta_{t-1}(v)/2}$. \\ 
Also, since by assumption $\W(T'(w_2,v)) \ge \W(T'(w_1,v)) $, we know that $\W(T'(w_2,v)) \ge
2^{\delta_{t-1}(v)/2} $. \\ 
 Further, since $\W(T'(z,v)) =  \W(T(z,v)) \ge \W(T'(w_1,v))$ we know that $\W(T'(z,v)) \ge
2^{\delta_{t-1}(v)/2}$. 
  
Hence,
 \begin{eqnarray*}
 \rem_t(v) & \ge & 2^{\delta_{t-1}(v)/2} + 2^{\delta_{t-1}(v)/2}\\
  & = & 2^{(\delta_{t-1}(v)+2)/2} \\
  & = & 2^{(\delta_{t}(v)+1)/2}
  \end{eqnarray*}

\end{enumerate}

\item \emph{$v$ is an internal node in $T'$}
\label{case: deg1-intnode}
\begin{figure}[h!]
\centering
\includegraphics[scale=0.3]{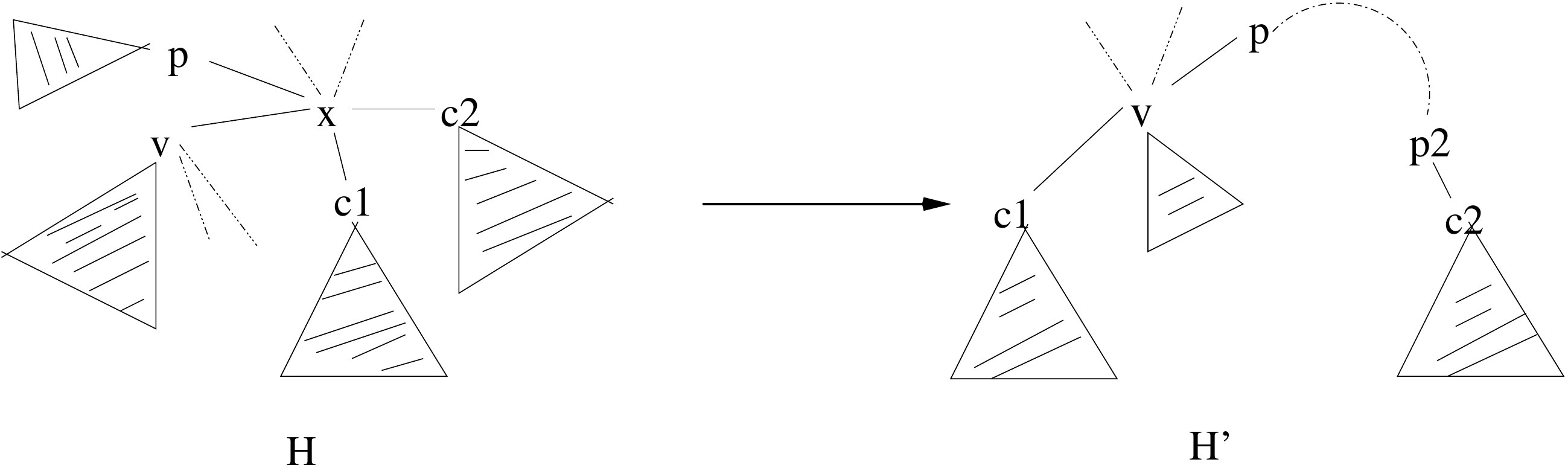}
\caption{Internal node $v$ with 1 child}
\label{fig: vinternalnode}
\end{figure}

\begin{figure}[h!]
\centering
\includegraphics[scale=0.3]{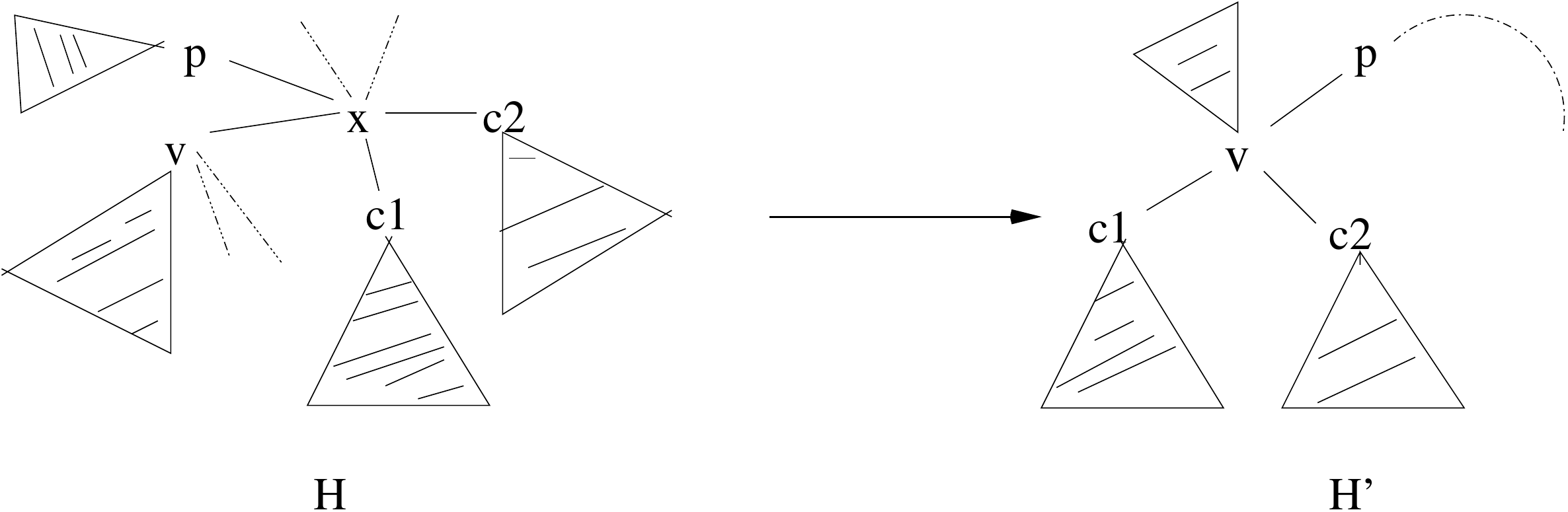}
\caption{Internal node $v$ with 2 children}
\label{fig: vdegplustwo}
\end{figure}

 For node $v$ to become an internal node, the deleted
neighbor $x$ must have at least three other neighbors.  Three neighbors of $x$  are shown as $C1$, $C2$ and $P$ in the
figures \ref{fig: vinternalnode} and \ref{fig: vdegplustwo}. 
Also, now $v$'s degree can increase by 1, as illustrated in figure \ref{fig: vinternalnode}, or by 2, as illustrated in figure  \ref{fig: vdegplustwo}.  Let us consider these cases separately:

\begin{enumerate}
 \item  $\delta_t(v) = \delta_{t-1}(v) + 1$

This can only happen when $v$ has a parent and a single child in $\RT$ as in figure \ref{fig: vinternalnode}. Let $P$ be
the parent of $v$ and
$C1$ the child of $v$.  $C1$ has to be a leaf node since the tree is complete and $v$ has only one child.  Observe that
there
exists at least one leaf node besides $C_{1}$ in the tree, accessible to $v$ only via $P$. Let this node be $C2$ and let
$P2$ be its parent. Note that $P2$ and $P$ may even be the same node. In our algorithm, any leaf node in $\RT$ has a
$\delta$ value no less than the $\delta$ value of any internal node. Thus,
 \begin{eqnarray*}
 \delta_{t-1}(C1) & \ge &\delta_{t-1}(v); \textrm{ and}\\
 \delta_{t-1}(C2) & \ge & \delta_{t-1}(v)
 \end{eqnarray*}

These inequalities, Lemmas \ref{lemma: nondecreasing} and \ref{lemma: WgreatRem}, and the  Inductive Hypothesis, imply
that
 \begin{eqnarray*}
\W(\T'(C1,v)) & \ge &  \rem_{t}(C1) \\
 &\ge & \rem_{t-1}(C1) \\
 &\ge & 2^{\delta_{t-1}(v)/2};\\
\W(\T'(C2,P2)) & \ge &  \rem_{t}(C2) \\
 &\ge & \rem_{t-1}(C2) \\
  &\ge & 2^{\delta_{t-1}(v)/2};\\
\W(\T(v, x)) & \ge & \rem_{t}(v)\\
  & \ge & \rem_{t-1}(v) \\
  &\ge &  2^{\delta_{t-1}(v)/2}.
 \end{eqnarray*}

 Since $\rem_t(v)$ can exclude at most one of $W(T'(C1,v))$, $W(T'(C2,P2))$ and $W(T(v, x))$, 
\begin{eqnarray*}
\rem_t(v) & \ge &  2^{\delta_{t-1}(v)/2} +  2^{\delta_{t-1}(v)/2} \\
  & = &  2^{(\delta_{t}(v) + 1)/2}
\end{eqnarray*} 

\item   $\delta_t(v) = \delta_{t-1}(v) + 2$
 
In this case $v$ has two children in $\RT$, $C1$ and $C2$, as illustrated in figure \ref{fig: vdegplustwo}. The analysis
is similar to the case above. The value  $\rem_t(v)$ can exclude at most one of $W(T'(C1,v))$, $W(T'(C2,v))$ and
$W(T(v,x))$ and we can show that all three of these values are at least  $2^{\delta_{t-1}(v)/2}$.
Thus, $ \rem_t(v)\ge 2^{(\delta_{t}(v))/2} $.
\end{enumerate}

\end{enumerate}

 Hence, the induction holds.
 
\end{proof}


\begin{lemma} For all vertices $v$, $rem(v)$ is always no more than n. \end{lemma}
\label{lemma: remNorless}
\begin{proof} No vertex is counted twice in a $rem$ value since the subtrees of a vertex are disjoint. Since the number of vertices in the subtrees cannot be more than the number of vertices remaining, the $rem$ value is always no more than the sum of the weights of all undeleted vertices in $G'$.

Define $W^*$ to be the sum of weights of all undeleted vertices in $G'$.  After initialization, $W^* = n$, since there are $n$ vertices.  At each step of the algorithm, $W^* = n$ , since the weight of the deleted vertex is added to one of the remaining vertices.  Thus, for node $v$,  $rem(v) \le n$. 

\end{proof}

\begin{lemma}
\label{l:degree} 
$\DASH$ increases the degree of any vertex by at most $O(\log n)$. \end{lemma}
\begin{proof} 

Every vertex $v$ starts with $\rem(v)=w(v)=1$. We know that $\rem(v)  \ge  2^{\delta(v)/2}$  \textrm{by Lemma \ref{lemma:
deglimit}}.  since $\rem(v)$ is at most n,  $2^{\delta(v)/2}  \le n$ . Taking $\log$ of both sides, $\delta(v)/2  \le 
\log n $.  Solving for $\delta(v)$ gives $ \delta(v)  \le  2 \log n  $. 

\end{proof}

\begin{lemma} 
\label{l:latency}
The latency to reconnect the network in $\DASH$ is $O(1)$.
\end{lemma}
\begin{proof}
During the reconnection process, $\DASH$ requires communication only between nodes one hop away, thus, the latency is just $O(1)$. 
\end{proof}

\begin{lemma} 
\label{l:messages}
The number of messages any node of initial degree $d$ sends out and receives is no more than $2(d + 2 \log n)\ln n $
\emph{with high probability} over all node deletions.
\end{lemma}
\begin{proof}
 In $\DASH$, after the reconnections have been made, messages are sent out by nodes when the minimum $ID$ has to be
propagated. With similarity to the \emph{record breaking problem} \cite{glick-rb78}, it is easily shown that \emph{w.h.p.}, a node
has its
$ID$ reduced no more than 2 $\ln n$ times, where the record is the node's $ID$. These are the only messages the node needs to 
 transmit or receive. Each time its $ID$ changes, the node sends this message to all its neighbors, Thus, it sends or receives
$O((d + \log n) \ln n)$ messages, since the final degree of the node is at most $d + 2 \log n$.

\end{proof}

\begin{lemma}
\label{l:latencyid}
 The amortized latency for  $ID$ propagation is $O(\log n)$ \emph{with high probability} over all node deletions.
\end{lemma}

\begin{proof}
Again, with similarity to the \emph{record breaking problem},  a node  sends messages to its neighbors
(neighbors, by definition, are a single hop away) only $O(\log n)$ times with high probability. Thus, messages are
transmitted $O(n \log n)$ times over all the nodes. Over $O(n)$ deletions, this implies that the amortized latency for
messages (involving $ID$ propagation) is only $O(\log n)$ .
\end{proof}
 
\SubSection{Proof of Theorem~\ref{theorem: DASH}}

The proof of Theorem~\ref{theorem: DASH} now follows immediately from Lemmas
\ref{l:degree}, \ref{l:latency}, \ref{l:messages} and \ref{l:latencyid}.
  


\section{Lower bounds on Locality-aware algorithms}
\label{sec: lower}


\SubSection{Necessity of Component tracking for healing strategies}
\label{subsec: comptrack}

To begin with, we give an insight as to why a healing strategy might need to keep track of connected components.

\begin{lemma}
\label{lemma: d+2} For  a tree, deletion of  a node of degree $d$ increases the sum total of degrees of its neighbors
by $d - 2$ for a locality-aware acyclic healing strategy.
\end{lemma}
\begin{proof}

A \emph{locality-aware acyclic healing strategy} will reconnect the neighbors of a deleted node without creating any cycles. If there were no cycles in the original graph involving the neighbors and not involving the deleted node, then such a strategy can only reconnect these neighbors as a tree to maintain their connectivity.

 A node of degree $d$ has $d$ neighbors. Since it was part of a tree, this node and its neighbors also constitute a tree. Let us call this the \emph{immediate subtree}. The \emph{immediate subtree} had $d$ edges and a total of $2d$ degrees. 
 These $d$ neighbors are now reconnected as a tree with $d-1$ edges and $2(d-1)$ degrees.
 Each of these neighbors lost a single degree due to the deletion of their edge to the deleted node.
 Thus, the total degrees gained on reconstruction are $2(d-1) - d = d - 2$. 
 
\end{proof}

 It is reasonable to assume that an efficient healing algorithm adds close to  the minimum possible edges at each step to maintain
connectivity of the neighbors of the deleted node. In $G'$, if a deleted node $v$ had two neighbors which had an
alternate path between themselves not involving $v$, then the algorithm may need to use only one of them for
reconnection to other nodes. By extension, if there were many neighbors which had alternate connections between them, the algorithm
may need to use only one of these nodes. This is equivalent to stating that the algorithm  may need to use only one node from a  connected component. Knowing that certain nodes are in the same component would allow the algorithm to do this.
 $G'$ is comprised only of edges added by the healing algorithm, and is always a forest. If the adversary mainly deletes nodes with degree greater than 2 and the algorithm does not use the component information, the sum total of degrees of the neighbors of the deleted nodes will increase by $ (d -2)$ i.e. at least 1, at each step. After many ($O(n)$) deletions,  only a few nodes will be left, and these will have $O(n)$ degree increase.

\subsection{A lower bound on healing by Degree-bounded locality-aware healing algorithms}
\label{subsec: loglognlowerbound}

 We now prove our result regarding the lower bounds for locality-aware algorithms in Theorem \ref{thm: lognbound}. Our lower bound occurs on graphs that are originally trees. To state the proof, we need to prove some other lemmas. 
  
First, we define the following operation that the adversary can perform on trees, where we assume self-healing is applied after every deletion:
\begin{description}
\item[Prune (r,s)]: For a node $r$ and its subtree headed by node $s$, the $Prune$ operation on $s$ leads to deletion of
all the nodes in that subtree including $s$. This operation can be accomplished by repeatedly deleting leaf nodes in the
subtree till all the nodes including $s$ are deleted. 

\begin{figure}[h!]
\centering
\includegraphics[scale=0.5]{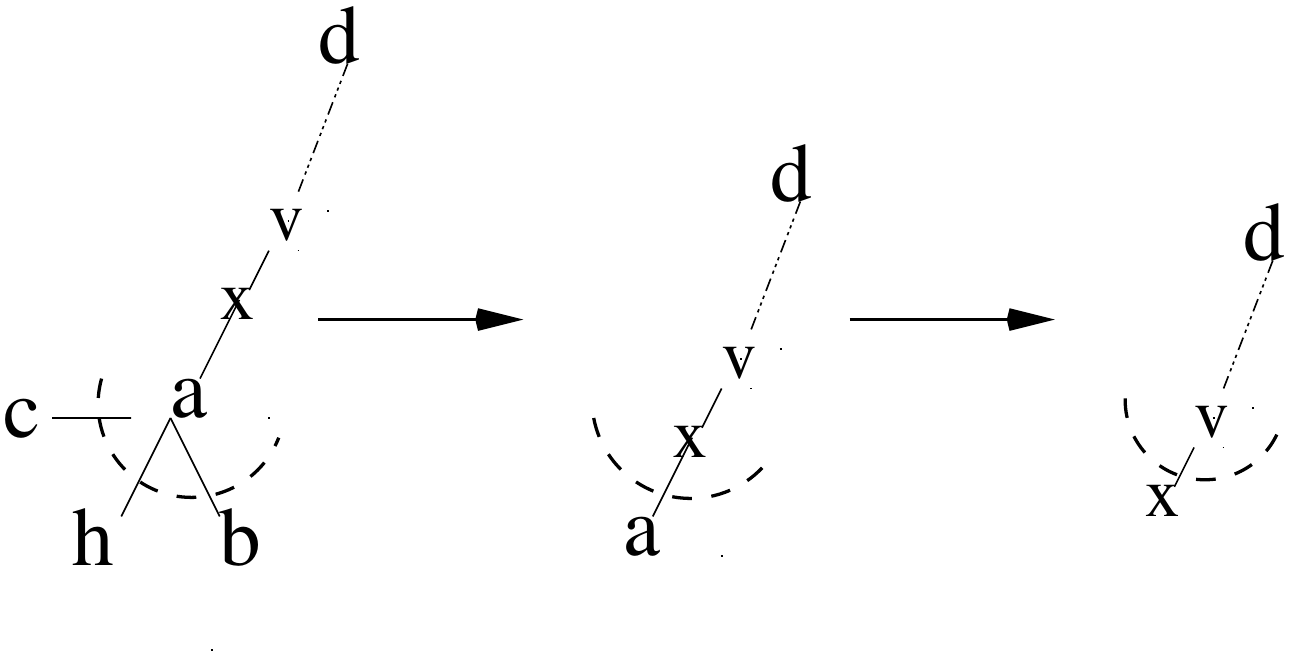}
\caption{Steps in Prune(v,x). Leaf nodes are deleted at each step. }
\label{fig: 3nbrs}
\end{figure}
\end{description}

\begin{lemma}
\label{lemma: degup} Deletion of a node with degree at least 3 increases the degree of at least one node by degree 1,
no matter how the healing occurs.
\end{lemma}
\begin{proof}

Any reconnection of more than two nodes has a 3-node line (as in figure \ref{fig: 3nbrs}) as a subgraph. Here the
internal node has a degree increase of $1$. Thus, at least one node increases it's degree by at least $1$.

\begin{figure}[h!]
\centering
\includegraphics{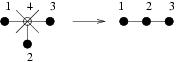}
\caption{An internal node  in  a 3-node line reconnection suffers a degree increase.}
\label{fig: 3nbrs}
\end{figure}

\end{proof}

For further discussion, we define the following:
\begin{description}
\item[Degree-bounded / M-degree-bounded :] A healing algorithm is \emph{degree-bounded} or \emph{M-degree-bounded} if
any node can increase its degree by at most $M$ in a single round of deletion and healing.
\end{description}


\begin{lemma}
\label{lemma: M+3up1} Consider a M-degree-bounded locality-aware healing algorithm used on a tree. In such a
situation, deletion of a node $v$ with degree at least M+3 leads to degree increase for at least two neighbors of $v$.
.\end{lemma}
\begin{proof}
 Node $v$ has $M+3$ neighbors. By Lemma \ref{lemma: d+2}, the sum total of degree increase of neighbors is $M+1$, when
the
graph is a tree. Since one node can get a maximum degree increase of $M$, at least one node has to incur the rest of the
degree increase. Thus, at least two nodes have to increase their degrees.

\end{proof}

\begin{figure}[h!]
\centering
\includegraphics[scale=0.5]{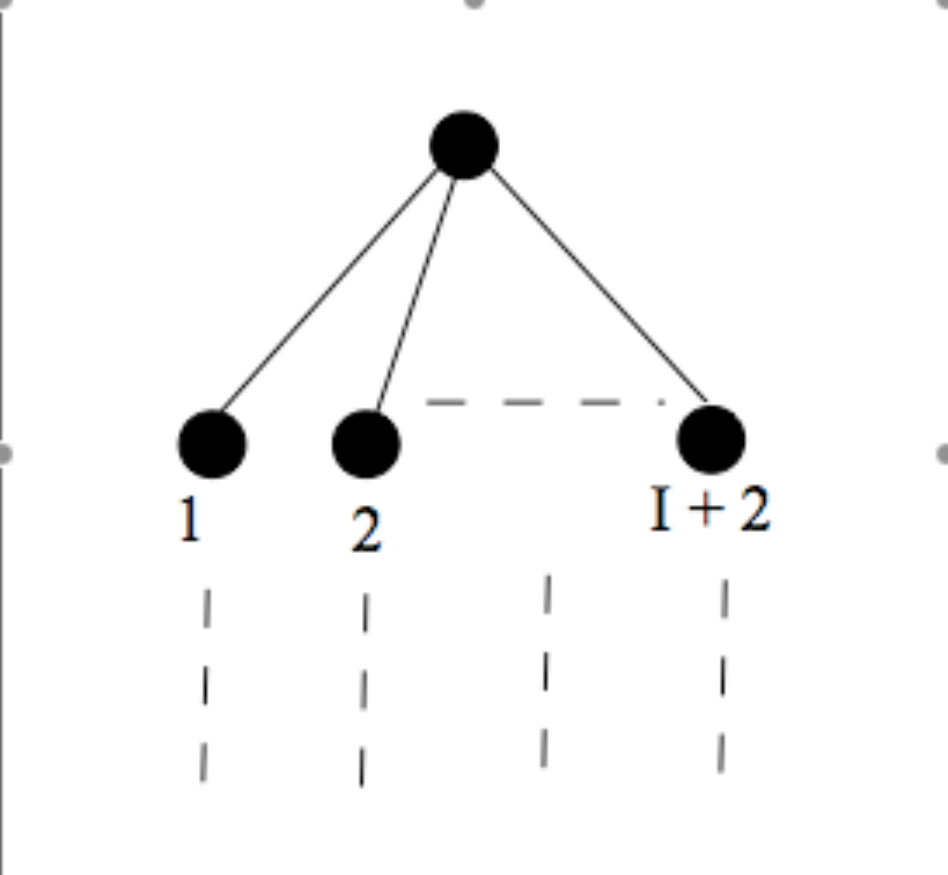}
\caption{M+2 -ary Tree}
\label{fig: Mplus2tree}
\end{figure}

\begin{algorithm}[h!]
\caption{\textsc{LevelAttack}: level-by-level attack on a (M+2)-ary tree}
\label{algo: levelattack}
\begin{algorithmic}[1]
\STMT Consider an (M+2)-ary tree $T$ of depth $D$ with levels numbered $0$ to $D$, the root being at level
$0$.
\STMT $i \leftarrow  D-1$
\WHILE {$i \ge 0$}
 \FOR{each node $v$ at level $i$} \label{levelalgo: forlevel}
 \STATE \label{levelalgo: prune} if $v$ has $c > M+2$ children  remove the excess $c - (M+2)$ nodes by deleting those
with least degree increases and their subtrees by using the \emph{Prune} operation, so that $v$ now has $M+2$ children.
\STATE delete $v$.
\ENDFOR
\STMT $i \leftarrow i-1$
\ENDWHILE
\end{algorithmic}
\end{algorithm}

Here, we introduce a new attack strategy:\\
\begin{description}
\item{\textsc{LevelAttack}:}
This strategy is described in Algorithm \ref{algo: levelattack}. In brief, the adversary deletes nodes one level at a
time beginning one level above the leaves of a $M+2$-ary complete tree going up to the root. 
The reasoning behind the strategy is the following: If the adversary deletes a node of degree $M+3$ in a tree, this
ensures that a degree increase of at least 1 is passed to its children. What the adversary must do is to  ensure that 
$log n$ of these degree increases are credited to the same node.
\end{description}

\begin{lemma}
\label{lemma: levelupD} Assume a $(M+2)-ary$ tree $T$, a degree-bounded locality-aware healing algorithm and the
\textsc{LevelAttack} adversarial strategy. Then, when \textsc{LevelAttack} deleted a node at level $i$,
$0 < i < D$ some leaf node of the original tree  increases its degree by at least $ D - i$.
\end{lemma}

\begin{proof} The proof is by induction.

\emph{Base case:} In the \textsc{LevelAttack} strategy, the nodes at level $D-1$ are deleted first. Thus, a deletion of a node at
$D-1$ is our base case. A node at level $D-1$ has $M+3$ neighbors. By lemma \ref{lemma:
M+3up1}, there is at least one leaf node that increases its degree by 1 or more. Thus, the base case
holds.

\emph{Inductive step:} Assume the hypothesis holds for nodes at level $i+1$. We now show that it holds for nodes at level $i$.
 Consider a node, say X at level $i \ge 0$ .  It had $M+2$ children at level $i+1$. By the inductive hypothesis, each of
these deletions led to at least one node with degree  $D-(i+1)$. Moreover, $X$ is not among  these $M+2$ nodes.
Moreover, all of  these are now neighbors of $X$, since $X$ itself was involved in each of these deletions.
 The \emph{Prune} algorithm in step \ref{levelalgo: prune} retains only these $M+2$ as children of $X$. Each of these children has degree increase $D-(i+1)$
and was originally a leaf node of $T$. The adversary now deletes $X$.  By lemma \ref{lemma: M+3up1}, at least one of these
children incurs a degree increase.


\end{proof}

\begin{theorem}
\label{thm: lognbound}
Consider any locality-aware algorithm that increases the degree of any node after an attack by at most a
fixed constant.  Then there exists a graph and a strategy of deletions on that graph that will force the algorithm to
increase the degree of some node  by at least $\log n$.
\end{theorem}

\begin{proof}
 
 It is sufficient to give a graph and an attack strategy such that any degree-bounded  locality-aware healing algorithm
will have to increase a particular node's degree by $\log n$. 
Let $M$ be the constant degree increase that is the maximum that the healing algorithm can impose on any one node in
the graph. Then, for a graph which is a full (M+2)-ary tree ( Figure \ref{fig: Mplus2tree}),
 the adversary uses  \textsc{LevelAttack}.  

 Consider a (M+2)-ary tree $T$ of depth $D$ with levels numbered
$0$ to $D$. By lemma \ref{lemma: levelupD},  after the last deletion in the
adversary strategy, which is the deletion of the  root of $T$ i.e. the node at level $0$ there is at least one node
left which has a degree increase of $D$. Since $D$ is $O(log n)$, this adversary strategy achieves a degree increase of
at least $O(log n)$.

\end{proof}

\Section{Experiments}
\label{sec: empirical}
 We carried out a number of experiments to ascertain the performance of various healing algorithms. We used a number of
attack strategies to measure how different healing strategies performed with regard to degree increase and stretch,
where stretch is the maximum ratio of distance increase in the healed network compared to the original network, over all
pairs of nodes. Our empirical results on stretch and a heuristic for maintaining low stretch
are described in Section \ref{app: stretch}. 
  
\SubSection{Methodology}
\label{subsec: methodology}
 Most of our experiments were
conducted  on random graphs. These graphs were generated by the \emph{Preferential attachment} model proposed by Barabasi \cite{barabasi-1999-286, barabasi-2003-50}.  The experimental approach was the following:
\begin{itemize}
\item For each graph size, for a particular deletion and healing strategy, repeat for 30 random instances of the graph:
\begin{itemize}
 \item  Repeat while there are nodes in the graph:
  \begin{itemize}
   \item delete a single node according to the deletion strategy.
   \item repair according to the self-healing strategy.
   \item measure the statistics (e.g. maximum change of degree for any node) for the graph. 
  \end{itemize}
  \end{itemize}
  \item average the statistics for each graph size.
 \end{itemize}
 
\SubSection{Attack Strategies}
\label{subsec: attack}

The aim of the adversary is to collapse the network by trying to overload a node beyond it's maximum capacity. There are
many possible attack strategies. One strategy is to delete the node with the maximum degree. We call this the $Max Node strategy$.
 It would seem that a strategy that leads to additional burden on an already high burden node would be a good strategy.
For the adversary,  one good adversarial strategy is to continuously attack/delete a randomly chosen neighbor of the
highest degree node in the network. We call this the $Neighbor of Max Strategy (NMS)$.
 This would also seem plausible as in a real network or the kind of networks we are looking at, it would be reasonable
that the hubs or the high degree nodes would be more well protected and resilient to attack while their less significant
neighbors should be easy to take down.

 \SubSection{Healing strategies}
\label{subsec: healing}
 We attempted various locality-aware healing strategies, some of which are the following:
 \begin{itemize}
 \item \emph{Graph heal}: On each deletion, we reconnect the neighbors of the deleted node in a binary tree regardless
of whether we introduced any cycles in the graph formed by the new edges introduced for healing. This seems to be  a
naive algorithm since the nodes use more edges than what are required for maintaining connectivity. 
 \item \emph{Binary tree heal}: On each deletion, we reconnect the neighbors of the deleted node in a binary tree being
careful not to introduce any cycles in the graph formed by the new edges introduced for healing. This is done using
random IDs which can then be used to identify which tree a particular node belongs to. This is an improvement on the
previous algorithm but still naive since it does not take into consideration the previous degree increase suffered by
nodes during healing.
 \item \emph{$\DASH$ (degree based binary tree heal)}:
 DASH is  smarter than the previous algorithms as borne out by the results of the experiments.  The DASH algorithm has been earlier described in Section \ref{subsec: dash} and stated as Algorithm \ref{algo: dash}.
\item \emph{$\SDASH$ (Surrogate degree based binary tree heal)}: (described in Section \ref{subsec: SDaSH})
 A heuristic based on $\DASH$ that tries to keep both node degrees and path lengths small.
 \end{itemize}

\SubSection{Degree increase}
\label{subsec: empirical-degree}

\begin{figure}[h!]
\centering
\includegraphics[width=3in, height=3in]{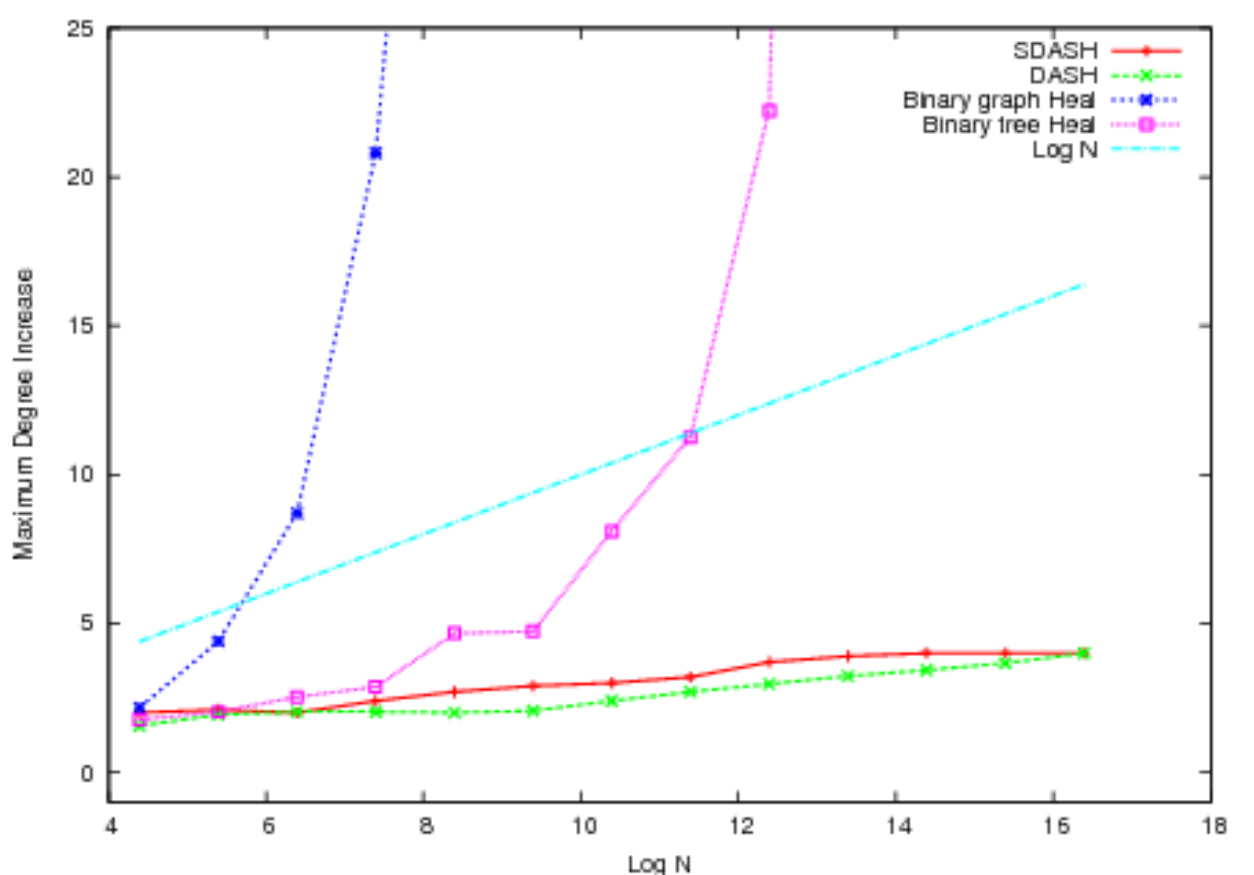}
\caption{Maximum Degree increase: DASH vs other algorithms}
\label{fig: Degplot}
\end{figure}

 The  $Neighbor of Max Strategy$ consistently resulted in higher degree increase, hence, we report results for only this attack strategy. Our experimental results clearly show that0 $\DASH$ and $\SDASH$  are good healing strategies. It performed well against both adversary strategies. Figure \ref{fig: Degplot} shows that $\DASH$ and $\SDASH$ have much lower degree increase than the other more naive strategies. Also, this degree increase was less than $\log n$, which is consistent with our theoretical results. $\SDASH$ has the additional nice property that it keeps path lengths small over multiple adversarial deletions.

\SubSection{Messages}
\label{subsec: empirical-messages}

\begin{figure}[h!]
   \centering
   \subfigure[ID changes for nodes]{
        \label{fig: nbrID}           
\includegraphics[width=3in, height=3in]{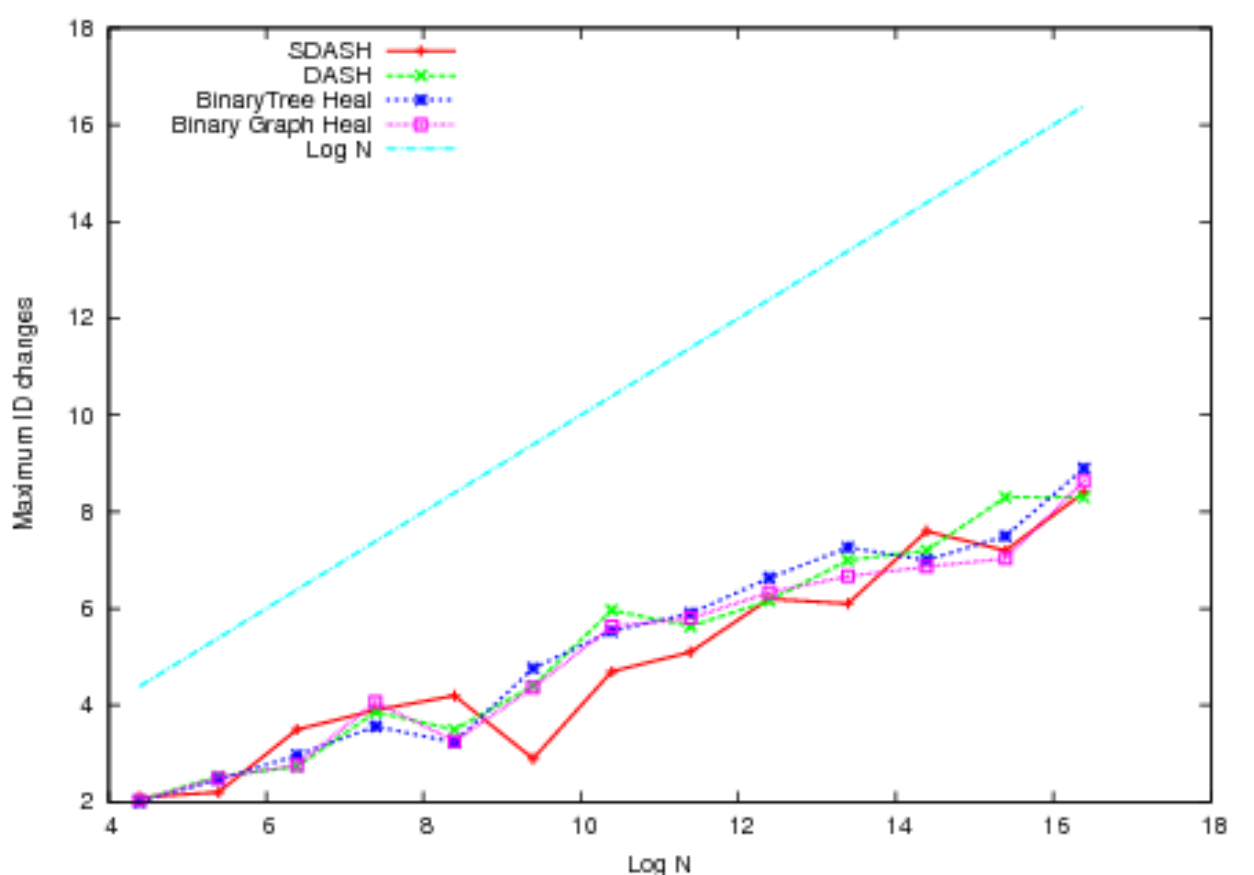}}
   \hspace{0.1in}
   \subfigure[Number of messages exchanged for Component(ID) information maintenance]{
        \label{fig: msgs}           
\includegraphics[width= 3in, height=3in]{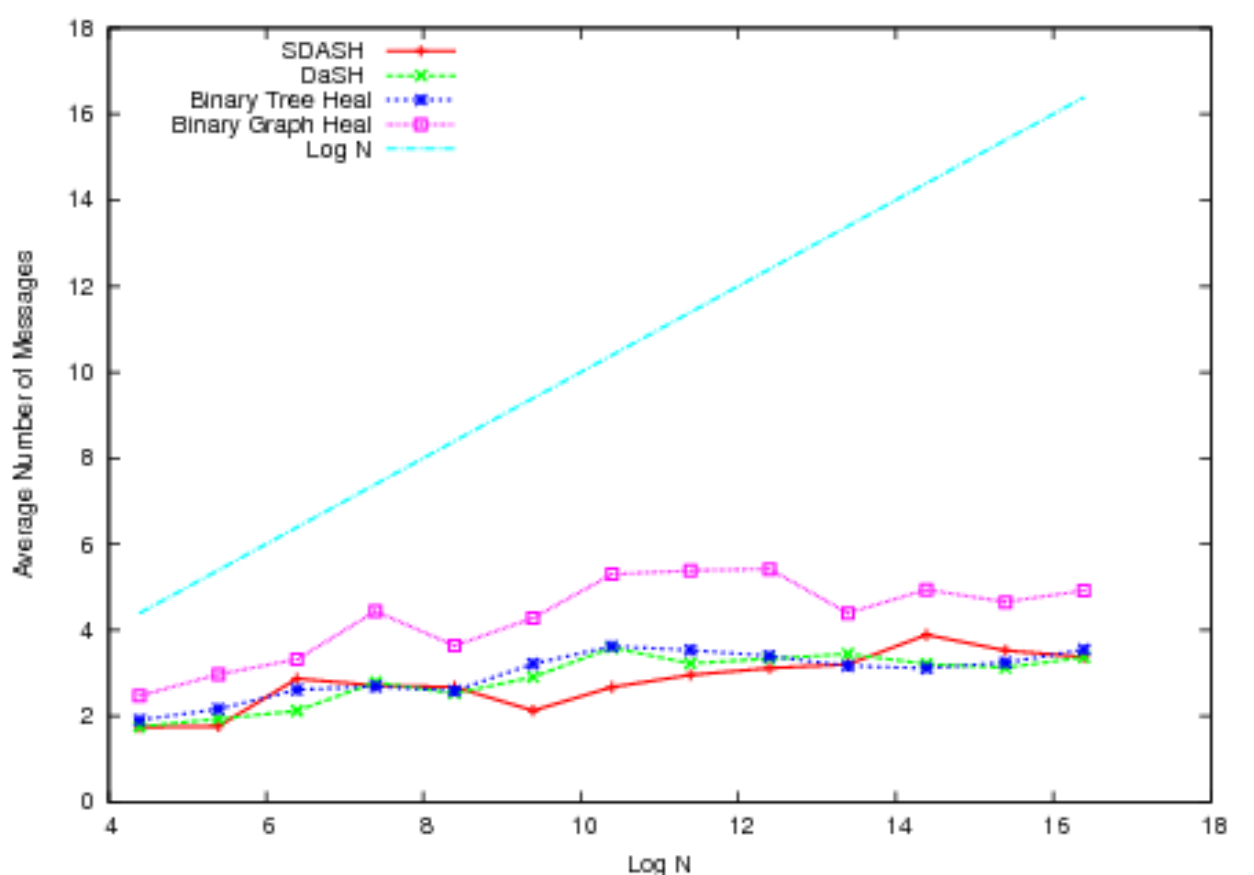}}
 \caption{ID changes and Messages exchanged per node}
   \label{fig: IDmsgs}                  
\end{figure}

Figure \ref{fig: nbrID} shows that the number of time a nodes $ID$ changes is less than $\log n$, as expected, for all healing strategies. Figure \ref{fig: msgs} shows the maximum number of messages a node sent out for the different strategies. Note that the number of messages a node sends out has to be less than or equal to the number of times a node changes ID times the degree of the node. Thus, algorithms with higher degree increase perform poorly.

\SubSection{Heuristics and experiments involving Stretch}
\label{app: stretch}

\subsubsection{Stretch}
Stretch is  an important property we would also like our self-healing algorithms to minimize. The stretch for
any two nodes is the ratio between their distance in the new healed network and their distance in the original network. Stretch for the network is the maximum stretch over all pairs of nodes. Stretch is also closely related to the diameter of the network. In some sense, maintaining low degree increase and low stretch are contradictory aims since a high-degree node will lead to shorter paths and possibly lower stretch in the network.

\subsubsection{$\SDASH$: a strategy with good empirical results}
\label{subsec: SDaSH}
 $\SDASH$ is an algorithm we have devised which empirically has both low degree increase and low stretch. During
self-healing, we say a node \emph{surrogates} if it replaces its deleted neighbor in the network. i.e. it takes all the
connections of the deleted neighbor to itself. Surrogation never increases stretch since the paths never increase in
length. In certain situations, it turns out that surrogation can be done without degree increase. In such situations,
$\SDASH$ does surrogation else it simply applies $\DASH$. $\SDASH$ is described in Algorithm \ref{algo: sdash}.

\begin{algorithm}[h!]
\caption{\textbf{SDASH:} Surrogate Degree-Based Self-Healing}
\label{algo: sdash}
\begin{algorithmic}[1]
\STATE \emph{Init:} for given network $G(V,E)$, Initialize each vertex with a random number $ID$ between [0,1] selected
uniformly at random. 
\WHILE {true}
\STATE \emph{If a vertex $v$ is deleted, do}
\STATE Let $m \in UN(v,G) \cup N(v,G')$ be the node with Maximum degree increase ($\delta$) of all nodes in $UN(v,G) \cup N(v,G')$.   
\IF {$ w \in UN(v,G) \cup N(v,G')$ and $\delta(w) + | UN(v,G) \cup N(v,G') | -1 \le \delta(m)$ }
\STATE connect all nodes in $ UN(v,G) \cup N(v,G')$ to $w$. 
\ELSE 
\STATE Nodes in $UN(v,G) \cup N(v,G')$ are reconnected into a \emph{complete binary tree}. To connect the tree, go left
to right, top down, mapping nodes to the \emph{complete binary tree} in increasing order of $\delta$ value.
\ENDIF
\STATE Let $MINID$ be the minimum $ID$ of any node in $UN(v,G) \cup N(v,G')$.
 Propagate $MINID$ to all the nodes in the tree of $UN(v,G) \cup N(v,G')$ in $G'$. All these nodes now set their $ID$ to
$MINID$.
\ENDWHILE
\end{algorithmic}
\end{algorithm}

 As can be seen in the figures that follow, $\SDASH$ seems to allow a degree increase  up to $O(\log n)$ and
stretch up to $O(\log n)$. We are working on proving theoretical properties of this algorithm.

\subsubsection{Stretch: empirical results}
\label{subsec: empirical-stretch}

\begin{figure}[h!]
\centering
\includegraphics[width=3in, height=3in]{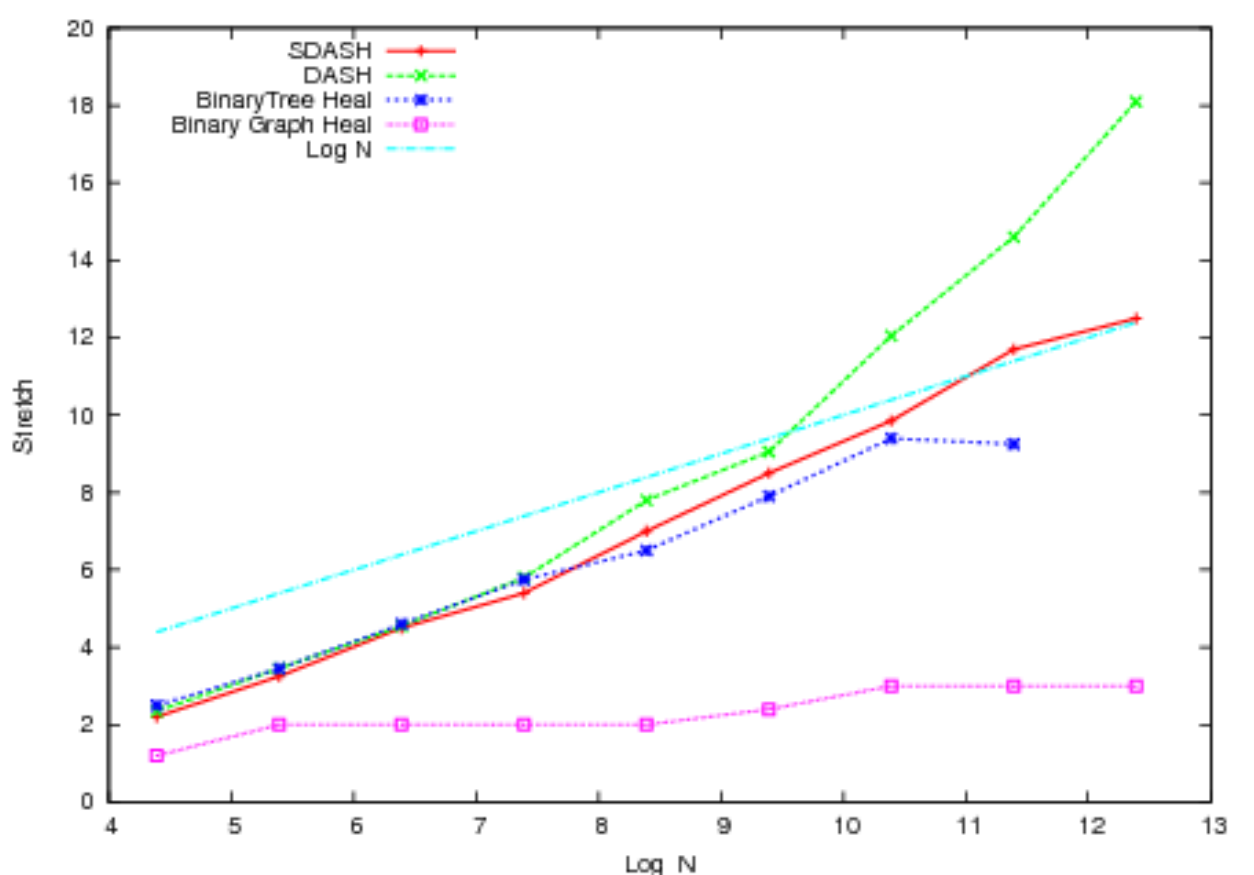}
\caption{Stretch for various algorithms}
\label{fig: stretchplot}
\end{figure}

 Figure \ref{fig: stretchplot} shows the performance of some of our algorithms for stretch.  We determined that the $Max Node strategy$ is most effective for the adversary when trying to maximize stretch and so our results in Figure \ref{fig: stretchplot} are against that adversarial strategy.   The more naive degree-control healing strategies do a good job of minimizing stretch.  However, it is important to keep in mind that these more naive algorithms increase the node degrees to a point where they are unlikely to be useful for many applications.  In contrast, our experiments show that $\SDASH$  does a good job of minimizing both stretch and degree increase.
  
\Section{Conclusions and future work}
\label{sec: conclusions}

We have studied the problem of self-healing in networks that are reconfigurable in the sense that new edges can be added to the network.  We have described $\DASH$, a simple, efficient and localized algorithm for self-healing, that provably maintains network connectivity, even while increasing the degree of any node by no more than $O(\log n)$.  We have shown that $\DASH$ is asymptotically optimal in terms of minimizing the degree increase of any node.  Further, we have presented empirical results on power-law networks showing that $\DASH$ significantly outperforms the naive algorithms for this problem.

Several interesting problems remain open including the following:  Can we not only maintain connectivity, but also provably ensure that lengths of shortest paths in the graph do not increase by too much?  Can we remove the need for propagating IDs in order to maintain connected component information, or is such information strictly necessary to keep the degree increase small?  Can we use the self-healing idea to protect invariants for combinatorial objects besides graphs?  For example, can we provide algorithms to rewire a circuit so that it maintains essential functionality even when multiple gates
fail?\\

\subsection{Acknowledgments}
 We gratefully acknowledge the help of Iching Boman, Dr. Deepak Kapur and his class \emph {Introduction to Proofs, Logic and Term-rewriting} and the UNM Computer Science Theory Seminar in writing this paper.  

\bibliographystyle{latex8}
\bibliography{selfhealltx8} 
\end{document}